\documentclass{llncs}%
\usepackage[pdftex]{graphicx}%
\usepackage{amsmath}%
\usepackage{amssymb}%
\usepackage{stmaryrd}%
\usepackage{url}%
\usepackage[all]{xy}%
\usepackage{caption}%
\usepackage{capt-of}%
\usepackage{paralist}%
\usepackage{color}%
\usepackage{enumitem}%
\usepackage[compact]{titlesec}%
\usepackage{algorithm}%
\usepackage{array}%
\newcommand{\pjs}[1]{\marginpar{\sc pjs}\textcolor{blue}{#1}}%
\newcommand{\ignore}[1]{}%
\newcommand{\lit}[1]{\left \llbracket #1 \right \rrbracket}
\newcommand{\true}{\mathit{true}}
\newcommand{\false}{\mathit{false}}
\newcommand{\wpmone}{\textsc{Wpm1}}
\newcommand{\msuone}{\textsc{Msu1}}
\newcommand{\msuthree}{\textsc{Msu3}}
\newcommand{\cumulative}{\texttt{cumulative}}
\DeclareMathOperator{\violated}{\mathit{violated}}

\def\mathrlap{\mathpalette\mathrlapinternal}

\def\mathrlapinternal#1#2{\rlap{$\mathsurround=0pt#1{#2}$}}

\newcolumntype{R}[1]{>{\raggedleft\arraybackslash}p{#1}}
\setlist{topsep=1pt plus 2pt}%
\setlist[trivlist]{topsep=6pt plus 2pt minus 1pt}%
\titlespacing*{\section} {0pt}{3ex plus .5ex minus .2ex}{2ex plus .2ex}%
\titlespacing*{\subsection} {0pt}{2.25ex plus .5ex minus .2ex}{1.5ex plus .2ex}%
\title{Unsatisfiable Cores for Constraint Programming}%
\author{Nicholas Downing \and Thibaut Feydy \and Peter J. Stuckey}%
\institute{%
National ICT Australia\thanks{NICTA is funded by the Australian Government as represented by the Department of Broadband, Communications and the Digital Economy and the Australian Research Council through the ICT Centre of Excellence program.}\ %
and the University of Melbourne, Victoria, Australia \\
\email{\{ndowning,tfeydy,pjs\}@csse.unimelb.edu.au}%
}%
\begin{document}%
%
\abovedisplayskip 6pt plus 2pt minus 1pt%
\abovedisplayshortskip 0pt plus 1pt%
\belowdisplayskip 6pt plus 2pt minus 1pt%
\belowdisplayshortskip 0pt plus 1pt%
\maketitle%
\begin{abstract}%
Constraint Programming (CP) solvers typically tackle optimization problems
by repeatedly finding solutions to a problem while placing tighter and
tighter bounds on the solution cost.  This approach is somewhat naive,
especially for soft-constraint optimization problems in which the soft
constraints are mostly satisfied.  Unsatisfiable-core approaches to solving
soft constraint problems in Boolean Satisfiability (e.g. MAXSAT) force all
soft constraints to hold initially. When solving fails they return an
unsatisfiable core, as a set of soft constraints that cannot hold
simultaneously.  Using this information the problem is relaxed to allow
certain soft constraint(s) to be violated and solving continues.  Since Lazy
Clause Generation (LCG) solvers can also return unsatisfiable cores we can
adapt the MAXSAT unsatisfiable core approach to CP.  We implement the
original MAXSAT unsatisfiable core solving algorithms \wpmone,  \msuthree{} 
in a state-of-the-art LCG solver and show that there exist 
problems which benefit from this hybrid approach.
\end{abstract}%
\section{Introduction}\label{sec:introduction}%
In this paper we consider how to make Constraint Programming (CP) solvers better at tackling soft-constraint problems.  CP solvers typically tackle optimization problems using branch-and-bound, and although soft-constraint problems can easily be mapped into a CP optimization framework, a distinguishing feature is that we expect most soft constraints to hold, at least on typical problems.  CP solvers rely heavily on propagation to cut down the search space, but soft constraints have little propagation ability (because even though the constraints are likely to be hold, we do not know for sure), and search takes over.

Hence, existing CP solvers are nearly always terrible at soft constraint
problems, a deficiency made worse by the fact that typical search strategies
are unaware of where the good solutions lie, such that in some cases,
thousands of solutions must be enumerated before the solver gets close to
proving optimality.  Indeed problems which are good for CP are almost 
always better with Mixed Integer Programming (MIP) 
once we soften the constraints. 
Here, we consider a different
approach, which is to lift the unsatisfiable-core solving approaches from
MAXSAT and integrate them into a CP solver.

By aggressively assuming that soft constraints hold, including the
intensional soft constraints characteristic of CP problems, we can either
find a solution or an \emph{unsatisfiable core}, as a set of soft
constraints which cannot hold simultaneously.  Given an unsatisfiable core
we adjust our assumptions and proceed, until feasibility is reached.  We
show that this approach maps well to CP solvers (as long as they can derive
an unsatisfiable core, which means in practice that the CP solver must use
Lazy Clause Generation), and that 
there are problems can benefit enormously from such an approach.

\section{Lazy Clause Generation (LCG)}\label{sec:lazyclausegeneration}%
We give a brief description of propagation-based solving and LCG,
for more details see~\cite{ohrimenko}.
We consider problems consisting
of constraints $\mathbf{C}$ over integer variables $x_1$, $\ldots$, $x_n$,
each with a given finite domain $D_\text{orig}(x_i)$.  
A feasible solution is a valuation $\theta$ to the variables, which
satisfies all constraints $\mathbf{C}$, and lies in the domain
$\mathbf{D}_\text{orig} = D_\text{orig}(x_1) \times \ldots \times
D_\text{orig}(x_n)$, i.e.~$\theta(x_i) \in D_\text{orig}(x_i)$.

A propagation solver keeps a domain restriction $D(x_i) \subseteq
D_\text{orig}(x_i)$ for each variable and considers only solutions that lie
within $\mathbf{D} = D(x_1) \times \ldots \times D(x_n)$.  Solving interleaves
propagation, which repeatedly applies propagators to remove unsupported
values, and search which splits the domain of some variable and considers
the resulting sub-problems.  This continues until all variables are fixed
(success) or failure is detected (backtrack and try another subproblem).

Lazy clause generation is implemented by introducing Boolean variables for
each potential value of a CP variable, named $\lit{x_i = j}$, and
for each bound, $\lit{x_i \ge j}$.  Negating them gives $\lit{x_i \ne j}$ and
$\lit{x_i \le j - 1}$.  Fixing such a \emph{literal} modifies $D(x_i)$ to
make the corresponding fact true, and vice versa. Hence the literals give an
alternate Boolean representation of the domain, which supports reasoning.
Lazy clause generation makes use of \emph{clauses} to record nogoods, where a
clause is a disjunction of (or essentially just a set of) literals.

In a lazy clause generation solver, the actions of propagators (and search)
to change domains are recorded in an \emph{implication graph} over the
literals.  Whenever a propagator changes a domain it must \emph{explain} how
the change occurred in terms of literals, that is, each literal $l$ that is
made true must be explained by a clause $L \rightarrow l$ where $L$ is a
conjunction of literals.  When the propagator detects failure it must
explain the failure as a \emph{nogood}, $L \rightarrow \textit{false}$, with
$L$ a conjunction of literals which cannot hold simultaneously.  
Then $L$ is used for conflict analysis~\cite{moskewicz} to generate a nogood
that explains the failure.

\section{MAXSAT solving algorithms}%
Before discussing how we integrate MAXSAT solving methods into CP, we
illustrate the original MAXSAT algorithms by means of a simple clausal
example.

\begin{example}\label{ex:maxsat}%
Consider Boolean variables $x_1$, $x_2$, $x_3$ and clausal constraints
\[
  C_1 \equiv x_1, \quad C_2 \equiv x_2, \quad C_3 \equiv x_3, \quad C_4 \equiv \neg x_1 \vee \neg x_2, \quad C_5 \equiv \neg x_1 \vee \neg x_3.
\]
Then $\{C_1, C_2, C_4\}$ is an unsatisfiable core, because $x_1 = x_2 =
\true$ violates $C_4$.  Similarly $\{C_1, C_3, C_5\}$ is an unsatisfiable
core, because $x_1 = x_3 = \true$ violates $C_5$.  As a Boolean
Satisfiability (SAT) problem this is infeasible.  As a Maximum
Satisfiability (MAXSAT) problem, we can allow some soft constraint(s) to be
violated.  For example if we relax $C_2$ and $C_3$ we find a solution $x_1 =
\true$, $x_2 = x_3 = \false$, with 2 violated constraints.
Or if we relax $C_1$ we find a better solution $x_1 = \false$,
$x_2 = x_3 = \true$, which has only one constraint violated.
\qed
\end{example}%
For MAXSAT we minimize $z = \sum_{j = 1}^n \violated(C_j)$ where $n$ is the
number of clauses and $\violated(C_j) =$ 0 if $C_j$ holds or 1 if $C_j$ is
violated.  A generalization is weighted partial MAXSAT where $z =
\sum_{j = 1}^n w_j \violated(C_j)$ given $w_j$ a weight attached to each
soft clause $C_j$, and hard clauses encoded by $w_j = \infty$.

\subsection{Branch-and-bound algorithm}\label{sec:branchandbound}%
For branch-and-bound we first convert the soft-constraint problem into a
hard-constraint problem using violator variables $v_j$.  We rewrite $C_j$
to 
$C_j \vee v_j$, that is, the original clause $C_j$ 
will now only be enforced if its violator $v_j$ is $\false$.

\begin{example}\label{ex:branchandbound}%
Adding violator variables $v_1, \ldots, v_5$ to the soft clauses $C_1, \ldots, C_5$ of Example~\ref{ex:maxsat} yields the hard-constraint optimization problem
\begin{align*}
  &\multicolumn{3}{l}{\text{minimize $z = v_1 + v_2 + v_3 + v_4 + v_5$ such that}}\\
  &C_1 \equiv x_1 \vee v_1, &\quad &C_2 \equiv x_2 \vee v_2, &\quad &C_3 \equiv x_3 \vee v_3,\\%
  &C_4 \equiv x_1 \vee \neg x_2 \vee v_4, &\quad& C_5 \equiv \neg x_1 \vee \neg x_3 \vee v_5.
\end{align*}
Violator variables which are $\false$ disappear, whereas violator variables which are $\true$ automatically satisfy their clause which hence plays no further role. \qed
\end{example}%
\begin{algorithm}%
  \setlength\topsep{0pt}%
  \begin{tabbing}%
    MM\=MM\=MM\=MM\=MM\=MM\=\kill%
    \textbf{inputs}: clauses $C_1, \ldots, C_n$ with weights $w_1, \ldots, w_n$ over variables $x_1, \ldots, x_m$\\%
    \textbf{outputs}: valuation $\theta_\text{opt}$ minimizing $z_\text{opt} =$ sum of weights of violated constraints\\%
    $(\theta_\text{opt}, z_\text{opt}) \leftarrow (\text{none}, \text{none})$\\%
    add violator variables to constraints:  $C_j \leftarrow C_j \cup \{v_j\}$ for $j \in 1..n$ where $w_j < \infty$\\%
    \textbf{while} SAT solver finds a valuation $\theta$ to the clause set \textbf{do}\\%
    \>$(\theta_\text{opt}, z_\text{opt}) \leftarrow \left(\theta, \sum_{j = 1}^n w_j \theta(v_j)\right)$, where $v_j =$ 1 for $\true$, 0 for $\false$/nonexistent\\%
    \>add a decomposition of the constraint $\sum_{j = 1}^n w_j v_j < z_\text{opt}$ to the clause set%
  \end{tabbing}%
  \caption{Branch-and-bound for weighted partial MAXSAT}\label{alg:branchandbound}%
\end{algorithm}%
Branch-and-bound search is defined in Algorithm~\ref{alg:branchandbound}.  We treat the problem as a hard-constraint satisfaction problem and simply find any solution $\theta$, calculate its objective value and add a new constraint to the problem enforcing that the next solution found should have an improved objective value.  When this fails, the most recent solution found (if any) is optimal.

The weakness of branch and bound for soft-constraint problems is that 
soft constraints do not propagate, so we need to set violator variables
$\false$ before the solver learns anything.  In contrast branch-and-bound
is very good for \emph{infeasible} problems since it detects infeasibility
in the first solve.

\subsection{Fu and Malik (\msuone{} or \wpmone{}) algorithm}\label{sec:wpmone}%
Fu and Malik~\cite{fu} proposed the \msuone{} algorithm for MAXSAT solving, later generalized by Ans\'otegui et al.~\cite{ansotegui} to \wpmone{} for the weighted case.  These algorithms iterate through a series of infeasible SAT problems until feasibility is reached.  When the SAT solver fails, it returns an unsatisfiable core as a set of clauses that cannot hold simultaneously.  Soft clauses in the set are relaxed using a fresh set of violator variables which are constrained so that at most one is $\true$ (\texttt{atmost1} constraint), and solving continues.  The first solution found is guaranteed to minimize the number of, or sum of weights of, violated clauses.

\begin{algorithm}%
  \setlength\topsep{0pt}%
  \begin{tabbing}%
    MM\=MM\=MM\=MM\=MM\=MM\=\kill%
    \textbf{inputs}: clauses $C_1, \ldots, C_n$ with weights $w_1, \ldots, w_n$ over variables $x_1, \ldots, x_m$\\%
    \textbf{outputs}: valuation $\theta_\text{opt}$ minimizing $z_\text{opt} =$ sum of weights of violated constraints\\%
    $z_\text{min} \leftarrow 0$\\%
    \textbf{repeat}\\%
    \>\textbf{if} SAT solver finds valuation $\theta$ to the clause set \textbf{then}\\%
    \>\>$(\theta_\text{opt}, z_\text{opt}) \leftarrow (\theta, z_\text{min})$; \textbf{break}\\%
    \>otherwise, SAT solver returns an unsatisfiable core $\{C_{u_1}, \ldots, C_{u_k}\}$\\%
    \>find minimum increase in $z$ implied by the core:  $w_\text{min} \leftarrow \min_{j = 1}^k w_{u_j}$\\%
    \>\textbf{if} $w_\text{min} = \infty$ \textbf{then}\\%
    \>\>$(\theta_\text{opt}, z_\text{opt}) \leftarrow (\text{none}, \text{none})$; \textbf{break}\\%
    \>$z_\text{min} \leftarrow z_\text{min} + w_\text{min}$\\%
    \>create a fresh set of violator variables we'll call $v_{u_1}, \ldots, v_{u_k}$ for now\\%
    \>\textbf{for} $j \in 1..k$ where $w_{u_j} < \infty$ \textbf{do}\\%
    \>\>\textbf{if} $w_{u_j} > w_\text{min}$ \textbf{then}\\%
    \>\>\>add a new copy of clause $C_{u_j}$ to the clause set with weight $w_{u_j} - w_\text{min}$\\%
    \>\>relax the original copy of the clause:  $(C_{u_j}, w_{u_j}) \leftarrow (C_{u_j} \cup \{v_{u_j}\}, w_\text{min})$\\%
    \>add a decomposition of the \texttt{atmost1} constraint $\sum_{j = 1}^k v_{u_j} \le 1$ to the clause set\\%
    \>delete all learnt clauses (or at least those invalidated by the above changes)%
  \end{tabbing}%
  \caption{\wpmone{} for weighted partial MAXSAT (\msuone{} is a special case)}\label{alg:wpmone}%
\end{algorithm}%

\wpmone{} or equivalently \msuone{} is defined in Algorithm~\ref{alg:wpmone}.  Solving the MAXSAT
problem as SAT with soft clauses considered hard, we find either a solution
or an unsatisfiable core.  In the latter case, we create a new MAXSAT
problem by encoding into it an allowance that we will not charge the first
$w_\text{min}$ units of the penalty for violating the clauses in the
unsatisfiable core.  If multiple clauses of the core are violated or if
violated clause(s) have weight greater than $w_\text{min}$ then the
remaining violation will be charged as usual.  The amount of penalty waived,
accumulates in $z_\text{min}$.  Eventually the MAXSAT problem is solved with
cost 0 (i.e.~all soft clauses are satisfied), then  
$z_\text{min}$ is the optimal solution cost.

Note the similarity of \wpmone{} with destructive lower bound search where
we set the objective $z = 0$, solve, if that fails increase it by one,
re-solve and repeat; the first solution found is optimal. \wpmone{} does
better by restricting where the violation is allowed to occur for $z = 1$
and at each further stage in the search.
%
\begin{example}\label{ex:wpmonemulti}%
To the problem of Example~\ref{ex:maxsat} we now add an extra variable $x_4$ and the extra clauses $C_6 \equiv x_4$ and $C_7 \equiv \neg x_3 \vee \neg x_4$.  The first unsatisfiable core is $\{C_1, C_3, C_5\}$.  Rewriting these clauses with violator variables $v_1$, $v_3$, $v_5$ gives
\[
  C_1 \equiv x_1 \vee v_1, \quad C_2 \equiv x_2, \quad C_3 \equiv x_3 \vee v_3, \quad C_4 \equiv \neg x_1 \vee \neg x_2, \quad C_5 \equiv \neg x_1 \vee \neg x_3 \vee v_5
\]
where $v_1 + v_3 + v_5 \le 1$ which we decompose to the additional hard clauses
\[
  \neg v_1 \vee \neg v_3, \quad \neg v_1 \vee \neg v_5, \quad \neg v_3 \vee \neg v_5.
\]
Then solving fails by deriving the empty clause as shown in Figure~\ref{fig:resolutiontree}.  The leaves of this tree show that the next core is $\{C_1, C_2, C_3, C_4, C_6, C_7\}$.  Relaxing the problem again (including $C_1$ and $C_3$ that had already been relaxed) gives\\[.1em]%
  \begin{minipage}[c]{.5\textwidth}%
    \begin{align*}%
      &C_1 \equiv x_1 \vee v_1 \vee v^\prime_1, &\: &C_2 \equiv x_2 \vee v^\prime_2,\\%
      &C_3 \equiv x_3 \vee v_3 \vee v^\prime_3, &\: &C_4 \equiv \neg x_1 \vee \neg x_2 \vee v^\prime_4,\\%
      &C_5 \equiv \neg x_1 \vee \neg x_3 \vee v_5, &\: &C_6 \equiv x_4 \vee v^\prime_6, \\
      &C_7 \equiv \neg x_3 \vee \neg x_4 \vee v^\prime_7, &\: &v_1 + v_3 + v_5 \le 1,\\%
      &\mathrlap{v^\prime_1 + v^\prime_2 + v^\prime_3 + v^\prime_4 + v^\prime_6 + v^\prime_7 \le 1.}
    \end{align*}\\[-.5em]%
This has a solution with $x_1 = x_2 = \true$, $x_3 = \false$, $x_4 = \true$,
which violates original clauses $C_3$ and $C_4$, and is optimal
with cost 2. \qed%
  \end{minipage}%
  \begin{minipage}[c]{.5\textwidth}%
    \centering \includegraphics{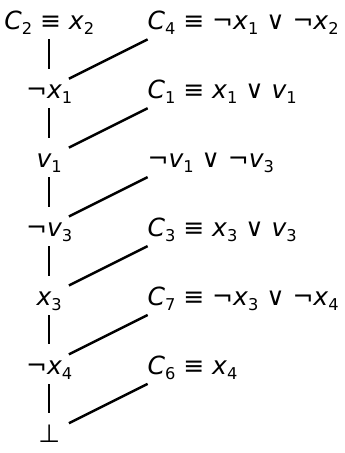}%
    \captionof{figure}{Example proof trace}\label{fig:resolutiontree}%
  \end{minipage}\\%
\end{example}%
\subsection{Marques-Silva \& Planes (\msuthree) algorithm}\label{sec:msuthree}%
The difficulty with \wpmone{} is that it is extremely aggressive, in the sense that the only soft-constraint violations allowed are those which are already known to exist.  On many problems the aggressive approach pays off, but on other problems either too many unsatisfiable cores need to be enumerated before achieving feasibility, or else the problems get more and more difficult to prove infeasible (the increasing number of \texttt{atmost1} constraints leads to an exponential number of relaxation-variable assignments, some of which may be symmetric).

\begin{algorithm}%
  \setlength\topsep{0pt}%
  \begin{tabbing}%
    MM\=MM\=MM\=MM\=MM\=MM\=\kill%
    \textbf{inputs}: clauses $C_1, \ldots, C_n$ with weights $w_1, \ldots, w_n$ over variables $x_1, \ldots, x_m$\\%
    \textbf{outputs}: valuation $\theta_\text{opt}$ minimizing $z_\text{opt} =$ sum of weights of violated constraints\\%
    $(\theta_\text{opt}, z_\text{opt}) \leftarrow (\text{none}, \text{none})$\\%
    \textbf{for} $j \in 1..n$ where $w_j < \infty$ \textbf{do}\\%
    \>add violator variable to constraint:  $C_j \leftarrow C_j \cup \{v_j\}$\\%
    \>add temporary singleton clause:  $C^\prime_j \leftarrow \{\neg v_j\}$\\%
    \textbf{repeat}\\%
    \>\textbf{if} SAT solver finds a valuation $\theta$ to the clause set \textbf{then}\\%
    \>\>$(\theta_\text{opt}, z_\text{opt}) \leftarrow \left(\theta, \sum_{j = 1}^n w_j \theta(v_j)\right)$, where $v_j$ is treated as in Algorithm~\ref{alg:branchandbound}\\%
    \>\>add a decomposition of the constraint $\sum_{j = 1}^n w_j v_j < z_\text{opt}$ to the clause set\\%
    \>\textbf{else}\\%
    \>\>SAT solver returns an unsatisfiable core $\{C_{u_1}, \ldots, C_{u_k}, C^\prime_{u^\prime_1}, \ldots, C^\prime_{u^\prime_\ell}\}$\\%
    \>\>\textbf{if} the unsatisfiable core contains no temporary clauses, i.e.~$\ell = 0$ \textbf{then}\\%
    \>\>\>\textbf{break}\\%
    \>\>delete the identified temporary clauses $C^\prime_{u^\prime_1}, \ldots, C^\prime_{u^\prime_\ell}$ from the problem\\%
    \>\>delete all learnt clauses (or at least those invalidated by the above change)%
  \end{tabbing}%
  \caption{\msuthree{} for weighted partial MAXSAT}\label{alg:msuthree}%
\end{algorithm}%
Marques-Silva \& Planes~\cite{marquessilva} 
proposed \msuthree{} for solving problems which
\wpmone{} does not handle efficiently for the above
reasons.  \msuthree{} is defined in
Algorithm~\ref{alg:msuthree}.  It is a hybrid unsatisfiable-core and
branch-and-bound approach, which leverages some of the benefits of
unsatisfiable-core solving, without being so aggressive.  All soft
constraints are considered hard initially, but each time an unsatisfiable
core is found, \emph{all} constraints in the set revert to soft, then the
ordinary branch-and-bound process continues, to minimize their violations.

\msuthree{} resembles binary search, we probe an initially overconstrained problem, if that fails we relax it, otherwise we constrain to find a better solution.

\begin{example}%
Rewriting Example~\ref{ex:maxsat} in the required format gives the same problem described in Example~\ref{ex:branchandbound} plus the additional temporary clauses
\[
  C^\prime_1 \equiv \neg v_1, \quad C^\prime_2 \equiv \neg v_2, \quad C^\prime_3 \equiv \neg v_3, \quad C^\prime_4 \equiv \neg v_4, \quad C^\prime_5 \equiv \neg v_5.
\]
Solving fails with unsatisfiable core $\{C_1, C_2, C_4, C^\prime_1, C^\prime_2, C^\prime_4\}$.  Removing $C^\prime_1$, $C^\prime_2$, $C^\prime_4$ from the problem yields the solution $x_1 = \false$, $x_2 = x_3 = \true$, $z = 1$.  Constraining $z \le 0$ propagates $v_1 = v_2 = v_4 = \false$ and returns the unsatisfiable core $\{C_1, C_2, C_4\}$, which has no temporaries, hence solving terminates. \qed
\end{example}%
\section{Unsatisfiable cores for LCG}

We can straightforwardly adapt the previously-described soft-constraint optimization approaches to CP.  A soft intensional constraint $I_j$ is represented as a half-reified constraint~\cite{feydy} of the form $i_j \rightarrow I_j$ where $i_j$ is the \emph{indicator variable} for the constraint $I_j$.  If $i_j$ is $\true$ then the constraint holds, and if $i_j$ is $\false$ then the constraint has no effect.

Note that a CP solver which has a propagator for the constraint $I_j$ can
straightforwardly be extended to provide a half-reified version of the
constraint.  Furthermore, the explanation algorithm for for $I_j$ in an 
LCG solver can also easily be extended to explain the half-reified version.

By adding indicator variables we effectively map the soft constraint problem to a MAXSAT problem.  If the soft intensional constraint $I_j$ has a weight $w_j$ then we add $C_j \equiv i_j$ as a soft, singleton, \emph{indicator clause} with weight $w_j$.  Now we can apply \wpmone{} or \msuthree{} effectively unchanged on the weighted indicator clauses.  Soft constraints are enforced when their indicator clauses hold.  Unsatisfiable cores enumerate conflicting indicator clauses and hence soft constraints.

For \wpmone{} (Algorithm~\ref{alg:wpmone}) these indicator clauses play an important role, as they will be progressively relaxed and won't necessarily be singletons by the end of solving.  For branch-and-bound (Algorithm~\ref{alg:branchandbound}) and \msuthree{} (Algorithm \ref{alg:msuthree}), the indicator clauses disappear (leaving only temporary clauses in the \msuthree{} case), because instead of augmenting the indicator clauses with violators $v_j$ and creating a useless implication, we can simply equate $i_j$ with $\neg v_j$.

To make use of soft global constraints that return a number of violations, we
can simply use literals encoding the integer violation count.
For example the constraint
 \texttt{soft\_alldifferent}($[x_1, \ldots, x_n], z)$ enforces that
$z$ is a violation count, e.g.~the number of pairs $x_i = x_j, 1 \leq i < j \leq n$. 
The usual LCG encoding of $z$ creates bounds variables
$\lit{z \geq 1}$, $\lit{z \geq 2}$, etc.
We can make an indicator clause from each of these literals 
with weights equal to the marginal cost of
each \texttt{soft\_alldifferent} violation, and hence map to a weighted soft clause problem.

\section{Experimental evaluation}%
To illustrate the potential usefulness of unsatisfiable core based optimization for CP, we consider a soft constraint variant of the Resource Constrained Project Scheduling Problems (RCPSP).  Rather than minimize makespan, we constrain the makespan to be some percentage of the optimal makespan, and soften all the precedences.  These problems are similar to RCPSP where the aim is to minimize the number of tardy jobs (that finish after their specified due date).

To create instances we take each RCPSP/max~\cite{bartusch} instance $M_i$ from the sets \textsf{ubo20}, \textsf{sm\_j30}, \textsf{ubo50} in PSPLib~\cite{kolisch}, which are systematically generated by ProGen/max~\cite{schwindt}, and a proven lower bound $l_i$ on its minimum makespan, usually the optimal makespan.  We constrain all tasks in $M_i$ to complete before time $\alpha l_i$ for each $\alpha \in \{.7, .8, .9\}$.  We maximize the number, or in a second experiment, the sum of randomly chosen weights $1..10$, of precedences that hold.


We aim to show that \wpmone{} and \msuthree{} can be advantageous over branch-and-bound, hence we run an LCG solver with all three methods.  A secondary aim is to show that LCG-based unsatisfiable core approaches can be superior to other solving technologies, so we provide best known decompositions to pseudo-Boolean (PB), MAXSAT and MIP, and evaluate them on \begin{inparaenum}[(i)]\item SAT-based PB solver MiniSAT+ 19/11/2012~\cite{een}, \item unsatisfiability-based MAXSAT solver MSUnCore 6/6/2011~\cite{marquessilva}, and \item MIP solvers CPLEX 12.4 and SCIP 3.0.1\end{inparaenum}.

For our own solver we used CPX, a state-of-the-art LCG solver, but modified to implement the \wpmone{} and \msuthree{} algorithms.  CPX and SCIP both use learning and a built-in \texttt{cumulative} propagator.  In these tests CPX uses activity-based search with phase saving and geometric restarts; the other solvers use their default searches which are similar, at least for the SAT-based solvers.

\ignore{
For the other solvers we encode the time decomposition of \cumulative{} (see e.g.~\cite{cumulative}), which propagates equivalently to the CP model, and is typically superior to other decompositions.  Start times and soft precedence constraints are implemented using the order encoding~\cite{tamura}, and resource constraints are either handled natively (MIP) or via the default encoding of MiniSAT+, which is typically decomposition to BDD (all SAT-based solvers).
}

We use a cluster of Dell PowerEdge 1950 with 2 $\times$ 2.0 GHz Intel Quad Core Xeon E5405, 2$\times$6MB Cache, 16 GB RAM, 600s timeouts, and 1GB memory limit per core.  Data files are available from \url{http://www.csse.unimelb.edu.au/~pjs/unsat_core}.  We disregard infeasible instances, where branch-and-bound will always be superior; almost all soft-constraint problems of interest are feasible.  We also disregard instances for which all solvers timed out.

\begin{table}[t]%
\caption{Comparative results for soft precedence RCPSP problems}\label{tab:res}%
  \centering%
  \scriptsize%
  \begin{tabular}{|l|R{7.67mm}R{2.67mm}|R{7.67mm}R{2.67mm}|R{7.67mm}R{2.67mm}|R{9mm}R{3mm}|R{9mm}R{3mm}|R{9mm}R{3mm}|R{9mm}R{3mm}|R{9mm}R{3mm}|}%
    \hline%
    \multicolumn{17}{|c|}{cardinality version}\\%
    \hline%
    $\alpha\;.7$\:\#ins&\multicolumn{2}{r|}{cpx b\&b}&\multicolumn{2}{r|}{cpx msu1}&\multicolumn{2}{r|}{cpx msu3}&\multicolumn{2}{r|}{sat b\&b}&\multicolumn{2}{r|}{sat msu1}&\multicolumn{2}{r|}{sat msu3}&\multicolumn{2}{r|}{cplex}&\multicolumn{2}{r|}{scip}\\
\hline
ubo20\hfill 53&0.386&\textbf{0}&0.142&4&\textbf{0.091}&\textbf{0}&1.046&\textbf{0}&1.083&1&1.312&\textbf{0}&16.350&7&120.897&16\\
j30\hfill 126&2.333&8&0.197&10&\textbf{0.188}&5&2.317&\textbf{0}&1.838&2&2.400&\textbf{0}&32.121&30&187.327&68\\
ubo50\hfill 47&16.543&6&\textbf{0.775}&4&0.894&\textbf{2}&515.777&36&349.598&20&475.797&33&174.902&21&531.704&44\\
\hline
$\alpha\;.8$\:\#ins&\multicolumn{2}{r|}{cpx b\&b}&\multicolumn{2}{r|}{cpx msu1}&\multicolumn{2}{r|}{cpx msu3}&\multicolumn{2}{r|}{sat b\&b}&\multicolumn{2}{r|}{sat msu1}&\multicolumn{2}{r|}{sat msu3}&\multicolumn{2}{r|}{cplex}&\multicolumn{2}{r|}{scip}\\
\hline
ubo20\hfill 61&0.236&\textbf{0}&\textbf{0.038}&1&\textbf{0.038}&\textbf{0}&0.933&\textbf{0}&0.773&1&1.072&\textbf{0}&9.059&7&77.723&14\\
j30\hfill 151&1.838&17&\textbf{0.097}&16&0.125&10&2.737&\textbf{1}&1.895&6&2.866&2&22.611&33&155.496&71\\
ubo50\hfill 56&9.668&7&\textbf{0.269}&1&0.398&\textbf{0}&446.866&39&314.035&24&458.802&39&178.648&25&523.275&51\\
\hline
$\alpha\;.9$\:\#ins&\multicolumn{2}{r|}{cpx b\&b}&\multicolumn{2}{r|}{cpx msu1}&\multicolumn{2}{r|}{cpx msu3}&\multicolumn{2}{r|}{sat b\&b}&\multicolumn{2}{r|}{sat msu1}&\multicolumn{2}{r|}{sat msu3}&\multicolumn{2}{r|}{cplex}&\multicolumn{2}{r|}{scip}\\
\hline
ubo20\hfill 64&0.124&\textbf{0}&\textbf{0.016}&1&0.022&\textbf{0}&0.865&\textbf{0}&0.598&\textbf{0}&0.937&\textbf{0}&6.727&4&61.616&10\\
j30\hfill 176&1.729&10&\textbf{0.057}&9&0.087&5&2.281&\textbf{1}&1.541&2&2.386&3&20.832&36&146.206&75\\
ubo50\hfill 63&6.540&3&\textbf{0.097}&\textbf{2}&0.227&\textbf{2}&385.716&41&275.759&29&380.389&46&161.992&23&504.252&56\\
\hline
\multicolumn{17}{|c|}{weighted version}\\
\hline
$\alpha\;.7$\:\#ins&\multicolumn{2}{r|}{cpx b\&b}&\multicolumn{2}{r|}{cpx wpm1}&\multicolumn{2}{r|}{cpx msu3}&\multicolumn{2}{r|}{sat b\&b}&\multicolumn{2}{r|}{sat wpm1}&\multicolumn{2}{r|}{sat msu3}&\multicolumn{2}{r|}{cplex}&\multicolumn{2}{r|}{scip}\\
\hline
ubo20\hfill 53&0.320&\textbf{0}&0.118&9&\textbf{0.091}&\textbf{0}&1.392&\textbf{0}&2.670&8&1.917&\textbf{0}&17.747&4&96.520&17\\
j30\hfill 128&1.420&6&\textbf{0.154}&21&0.181&2&3.537&\textbf{1}&3.236&12&3.298&\textbf{1}&29.061&34&176.513&71\\
ubo50\hfill 48&10.783&3&\textbf{0.548}&8&0.871&\textbf{1}&549.042&40&329.905&19&501.536&37&168.355&24&516.557&44\\
\hline
$\alpha\;.8$\:\#ins&\multicolumn{2}{r|}{cpx b\&b}&\multicolumn{2}{r|}{cpx wpm1}&\multicolumn{2}{r|}{cpx msu3}&\multicolumn{2}{r|}{sat b\&b}&\multicolumn{2}{r|}{sat wpm1}&\multicolumn{2}{r|}{sat msu3}&\multicolumn{2}{r|}{cplex}&\multicolumn{2}{r|}{scip}\\
\hline
ubo20\hfill 61&0.196&\textbf{0}&0.058&6&\textbf{0.048}&\textbf{0}&1.040&\textbf{0}&1.080&3&1.140&\textbf{0}&7.813&5&59.778&14\\
j30\hfill 151&1.250&8&\textbf{0.081}&16&0.137&8&3.349&\textbf{1}&2.556&15&2.987&2&21.072&31&147.957&69\\
ubo50\hfill 56&9.252&2&\textbf{0.167}&2&0.449&\textbf{0}&476.299&39&300.131&20&439.167&38&129.856&22&455.111&46\\
\hline
$\alpha\;.9$\:\#ins&\multicolumn{2}{r|}{cpx b\&b}&\multicolumn{2}{r|}{cpx wpm1}&\multicolumn{2}{r|}{cpx msu3}&\multicolumn{2}{r|}{sat b\&b}&\multicolumn{2}{r|}{sat wpm1}&\multicolumn{2}{r|}{sat msu3}&\multicolumn{2}{r|}{cplex}&\multicolumn{2}{r|}{scip}\\
\hline
ubo20\hfill 64&0.137&\textbf{0}&\textbf{0.015}&2&0.027&\textbf{0}&1.015&\textbf{0}&0.678&2&0.940&\textbf{0}&4.967&4&36.493&7\\
j30\hfill 177&1.206&5&\textbf{0.060}&14&0.109&4&2.811&\textbf{1}&1.912&11&2.376&\textbf{1}&17.888&40&128.818&74\\
ubo50\hfill 62&6.739&2&\textbf{0.077}&2&0.314&\textbf{0}&411.847&44&232.487&20&373.572&40&102.812&22&438.892&54\\
    \hline%
  \end{tabular}%
\end{table}%
%
The results shown in Table~\ref{tab:res} compare our solver CPX using branch-and-bound, \wpmone{} (or \msuone{} as a special case) and \msuthree{} (first three columns) with the SAT-based solvers MiniSAT+ using branch-and-bound and MSUnCore using \wpmone{} or \msuthree{} (next three columns) and the MIP solvers (last two columns).  The numbers shown are geometric mean of solving time (s; using 600s as solving time for instances that timed out), followed by number of timeouts.  The best solving time and (equal-)best number of timeouts are highlighted.

The results show that both \wpmone{} and \msuthree{} can be \emph{highly advantageous} over branch-and-bound when used with a learning CP solver.  As the makespan constraint becomes more generous, the percentage of soft constraints that can hold increases, and the advantages of \wpmone{} and \msuthree{} over branch-and-bound (in assuming that soft constraints hold), become more pronounced.

As the problems become larger, the huge numbers of variables created by the decompositions begin to overwhelm the SAT/MIP solvers, demonstrating the importance of using an LCG solving approach \emph{in addition to} the unsatisfiability-based algorithms \wpmone{} and \msuthree{} already available in the MSUnCore solver.

\section{Related work and conclusion}%

Specialized solvers~\cite{degivry,larrosa} have been highly successful for soft-constraint CSPs 
in \emph{extensional form}.  These approaches are similar to \wpmone{}
as they both effectively shift part of the cost function as inconsistencies
are detected.  But many problems (such as the scheduling problem we
investigate) 
are not feasible to encode using
extensional constraints only. We are unaware of any other 
approaches to soft intensionally defined
constraint problems beyond branch-and-bound, apart from
PBO/WBO~\cite{een,manquinho} which support intensionally-defined \textit{linear}
constraints only.
 
In this paper we demonstrate how to use unsatisfiable-core methods developed
for MAXSAT to solve CP optimization problems containing soft constraints,
by making use of the facility of LCG solvers to generate unsatisfiable
cores.
The results clearly show that CP solvers should incorporate unsatisfiable
core optimization algorithms, since they can be dramatically superior to
branch-and-bound on appropriate problems. 
\ignore{
Future work is to extend the use of unsatisfiable core approaches to soft global constraints, and to objectives based on integer rather than only Boolean variables (for example, preferences).
}

\bibliography{unsat_core}
\bibliographystyle{splncs03}
\end{document}